\documentclass[12pt]{article}

\usepackage[pdftex]{graphicx}
\usepackage[pdftex]{color}

\usepackage{epsfig,longtable}
\usepackage{makeidx}
\usepackage{varioref}
\usepackage{xr}
\usepackage{amsmath}
\usepackage{amsthm}

\usepackage{amssymb}
\usepackage{url}
\usepackage[authoryear]{natbib}
\usepackage{setspace}
\usepackage{subfigure}
\usepackage{multirow}
\usepackage{dsfont}
\newtheorem{definition}{Definition}[section]

\numberwithin{equation}{section}

\begin{document}
 
\title{Computing the Bergsma Dassios sign-covariance}
\maketitle

\begin{center}
Yair Heller \\
\emph{ E-mail: heller.yair@gmail.com }\\
Ruth Heller \\
\emph{Department of Statistics and Operations Research, Tel-Aviv
university, Tel-Aviv, Israel. E-mail: ruheller@post.tau.ac.il}\\
\end{center}

\begin{abstract}
\cite{Bergsma} introduced an independence measure which is zero if and only if two random variables are independent. This measure can be naively calculated in $O(n^4)$. \cite{Weihs} showed that it can be calculated in $O(n^2 \log n)$. In this note we will show that using the methods described in \cite{Hellera}, the measure can easily be calculated in only $O(n^2)$.  \
\end{abstract}

\section{Introduction}
Testing whether two ordinal random variables are independent given a sample  ${(x_i, y_i)}_{i=1}^n$ is a classic problem in statistics. Early efforts such as Pearson's correlation and Kendall's $\tau$ focused on testing against linear or monotone relationships. The first test for any type of independence was provided by \cite{Hoeffding}. This test is based on multiple partitions of the $(X,Y)$ plane into four quadrants where the number of points in a quadrant is  compared to what it would be under independence. This score can be calculated in $O(n \log n)$ by counting the inversions of the permutation from the ranks of  ${(x_i)}_{i=1}^n$ to ${(y_i)}_{i=1}^n$ as described in \cite{Hellerb}. Additional rank based scores were suggested by others, typically based on finer partitions of the plane (see \cite{Hellera} for review, algorithms and a powerful method that takes into account all possible partitions).

\cite{Bergsma} present another measure of independence which in the case of no ties is, also based on partitions of the plane into four quadrants. This method is in some sense a generalization of Kendall's $\tau$. Using the notation of \cite{Weihs} (in their equation 1) the statistic is defined as:
\begin{align}
\label{eq:t_star}
	t^* &:= \frac{(n-4)!}{n!}\sum_{\substack{1\leq i,j,k,l \leq n \\ i,j,k,l\ \text{distinct}}} a(x_i,x_j,x_k,x_l)a(y_i,y_j,y_k,y_l) 
\end{align}
where
\begin{align}
	a(z_1,z_2,z_3,z_4) := sign(|z_1-z_2| + |z_3-z_4| - |z_1-z_3| - |z_2-z_4|).
\end{align}
This definition clearly shows that the statistic can be naively calculated in $O(n^4)$ since one can simply go over all quadruples of points. However, \cite{Weihs} show that the statistic can be calculated in $O(n^2logn)$ using red black trees. In this note we will show that the statistic can in fact be calculated in only $O(n^2)$ using methods described in \cite{Hellera}.

\section{The algorithms}

As \cite{Bergsma} show, the score is actually based on the number of concordant quadruples vs. the number of discordant quadruples. In a manner analogous to Kendall's $\tau$ concordance and discordance are defined for a quadruple ${(x_i, y_i)}_{i=1}^4$ as follows:
\begin{definition}
A quadruple is called \textbf{concordant} if either  ($\max(x_1,x_2)<\min(x_3,x_4)$ and $\max(y_1,y_2)<\min(y_3,y_4)$) or ($\max(x_1,x_2)<\min(x_3,x_4)$ and $\min(y_1,y_2)>\max(y_3,y_4)$)
\end{definition}
\begin{definition}
A quadruple is called \textbf{discordant} if $\max(x_1,x_2)<\min(x_3,x_4)$ and $\max(y_1,y_2)>\min(y_3,y_4)$ and $\min(y_1,y_2)<\max(y_3,y_4)$
\end{definition}
We start with the simple case without ties.

\subsection{The algorithm without ties}
It is easy to see that in this case the statistic reduces to (equation 4 in \cite{Weihs})
\begin{align*}
	t^* =  \frac{(n-4)!}{n!}(24\cdot N_c) - \frac{1}{3}, \label{eq:simple_untied_t_star}
\end{align*}
where $N_c$ is the number of concordant quadruples. Therefore all we need to do is calculate $N_c$. Again as in \cite{Weihs} et. al clearly:

\begin{align*}
	N_c = \sum_{3\leq k\leq n-1} \sum_{k<l\leq n}{M_<(k,l) \choose 2} + {M_>(k,l) \choose 2}
\end{align*}
where they define
\begin{align*}
	M_<(k,l) &:= |\{i:\ x_i < \min(x_k, x_l),\ y_i < \min(y_k, y_l)\}|, \\
	M_>(k,l) &:= |\{i:\ x_i < \min(x_k, x_l),\ y_i > \max(y_k, y_l)\}|.
\end{align*}
However contrary to their algorithm we show that $M_<(k,l)$ and $M_>(k,l)$ can be calculated in $O(1)$ and not in $O(\log n)$ after a preprocessing step which takes $O(n^2)$. This can be done by the methods described in \cite{Hellera} as follows:
We first note that the statistic is only based on ranks so we transform every pair $(x_i,y_i)$ to its respective ranks $(r_i,s_i)$ where $r_i,s_i\in\{1,...,n\}$, this can of course be done in $O(n \log n)$.
We can now calculate the empirical cumulative distribution 
\begin{equation}\label{eq-agg-sum-ind-A}
A(r, s) = \sum_{i=1}^nI(r_i\leq r \ \textrm{and} \  s_i\leq s), \quad (r,s) \in \{0,1,\ldots,n \}^2 
\end{equation}
(where $A(0,s) =0, A(r,0)=0$) in $O(n^2)$ time and space:

First, let $B$ be the $(n+1)\times (n+1)$  zero matrix, and initialize to one $B(r_i, s_i)$ for each observation $i=1,\ldots,n$. Next, go over the grid in $s$-major order, i.e., for every $s$ go over all values of $r$, and compute:
\begin{enumerate}
\item $A(r, s) = B(r , s - 1) + B(r-1, s ) - B(r - 1, s - 1) + B(r, s)$, and
\item $B(r,s) = A(r,s)$.
\end{enumerate}
It is easy to see that $M_<(k,l)=A[rank(min(x_k,x_l))-1,rank(min(y_k,y_l))-1]$ and similarly that $M_>(k,l)=rank(min(x_k,x_l))-A[rank(min(x_k,x_l)),rank(max(y_k,y_l))]$ and therefore for each $k,l$ $M_<(k,l)$ and $M_>(k,l)$ can be calculated in O(1), resulting in a total of $O(n^2)$ as desired.

\subsection{The algorithm for data with ties}
First, for ease of notation we order the samples such that $x_1 \leq x_2...\leq x_n$.
By Lemma 1 in \cite{Weihs} in this case the score reduces to
\begin{align*}
	t^* &=  \frac{(n-4)!}{n!}(16\cdot N_c - 8\cdot N_d) \\
\end{align*}

Therefore, greater care must be taken in this case as it requires calculating also $N_d$ (where $N_d$ is the number of discordant quadruples), which with ties is a little more subtle since a quadruple can be neither concordant nor discordant. We will use the following ranking scheme - $n$ observations with $m$ unique values will be transformed to $n$ ranks in the range $1...m$ (so for example ${2,2,3.5,4,4,4}$ will be ranked as ${1,1,2,3,3,3}$). We first note that calculating the number of concordant pairs can be done in the same way as in the section above without ties, except that  when we calculate the empirical cumulative distribution $B(r, s)$ will be initialized to the \textbf{number} of observations with ranks $(r,s)$, which can be greater than one. 
We now turn to computing $N_d$. Define $N_d(k,l)=|\{i<j<k : i,j,k,l \text { are discordant}\}|$. Clearly $N_d= \sum_{3\leq k\leq n-1} \sum_{k<l\leq n}{N_d(k,l)}$. Following \cite{Weihs} for any pair of samples $(x_k,y_k)$ and $(x_l,y_l)$ such that $k<l$ we define: 
\begin{allowdisplaybreaks}
\begin{align*}
	top(k,l) &= |\{i: x_i<x_k\ \text{and}\ y_i> \max(y_k,y_l)\} | ,   \\
	mid(k,l) &= |\{i: x_i<x_k\ \text{and} \ \min(y_k,y_l) < y_i< \max(y_k,y_l)\} | ,\\
	bot(k,l) &= |\{i: x_i<x_k\ \text{and}\ y_i< \min(y_k,y_l)\} | ,\\
	\mathit{eqMin}(k,l) &= |\{i: x_i<x_k\ \text{and}\ y_i = \min(y_k,y_l)\}  |,\\
	\mathit{eqMax}(k,l) &= |\{i: x_i<x_k\ \text{and}\ y_i = \max(y_k,y_l)\} |. 
\end{align*}
\end{allowdisplaybreaks}
Quoting equations 11 and 12 in \cite{Weihs}
if $y_k=y_l$ then 
\begin{align*}
	N_d(k,l) = 0,
\end{align*}
and if $y_k\not =y_l$ then 

\begin{equation} \label{eq:N_d}
\begin{aligned}
	N_d(k,l) &= top(k,l)\cdot \left(mid(k,l) + \mathit{eqMin}(k,l)+ bot(k,l)\right)  \\
	&+ bot(k,l)\cdot \left(mid(k,l) + \mathit{eqMax}(k,l)\right) \\
	&+ \mathit{eqMin}(k,l)\cdot (mid(k,l) + \mathit{eqMax}(k,l)) \\
	&+ \mathit{eqMax}(k,l)\cdot mid(k,l) \\
	& + {mid(k,l) \choose 2} - \sum_{y\in unique(k,l)}{|\{1\leq i< k: x_k\not= x_i\ \text{and}\ y_i=y\}| \choose 2} 
\end{aligned}
\end{equation}
where 
\begin{align*}
	unique(k,l) := \{y_i : 1\leq i< k\ \text{and}\ x_i\not=x_k\ \text{and}\ \min(y_k,y_l) <y_i < \max(y_k,y_l)\}.
\end{align*}
Clearly, $top(k,l)$, $mid(k,l)$, $bot(k,l)$, $eqMin(k,l)$,$eqMax(k,l)$ can be calculated using the empirical cumulative distribution in $O(1)$ as described in the previous section (e.g. $mid(k,l)= A[rank(x_k)-1,rank(max(y_k,y_l))-1]-A[rank(x_k)-1,rank(min(y_k,y_l))]$ and $eqMin(k,l)=A[rank(x_k)-1,rank(min(y_k,y_l))]-A[rank(x_k)-1,rank(min(y_k,y_l))-1]$).

We will now show how to calculate the last element in equation \ref{eq:N_d}. This will be done with a procedure similar to the one used to calculate the empirical cumulative distribution in the previous section. Our first goal will be to calculate in $O(n^2)$

\begin{align*}
A(r,s)=\sum_{y\  s.t. \ rank(y)<=s}{|\{i: rank(x_i)<r \ \text{and}\ y_i=y\}| \choose 2}
\end{align*}

We initialize $A(0,s)=A(r,0)=0$. We further set $B(r,s)=|\{i:rank(x_i)=r \text{ and } rank(y_i)=s\}|$
We now compute the cumulative row sum $R(r,s)=R(r-1,s)+B(r,s)$ and then we compute row by row $A(r,s)=A(r,s-1)+{R(r,s) \choose 2}$.
Once we have $A(r,s)$ we can easily calculate the last element in \ref{eq:N_d} in O(1).

\begin{align*}
\sum_{y\in unique(k,l)}{|\{1\leq i< k: x_k\not= x_i\ \text{and}\ y_i=y\}| \choose 2}=A[rank(x_k)-1,rank(max(y_k,y_l))-1]- \\ 
A[rank(x_k)-1,rank(min(y_k,y_l))]
\end{align*}

Thus completing the computation in $O(n^2)$ as required.

\section{Conclusion}
We have shown how to calculate the Bergsma Dassios association measure in $O(n^2)$. However, the question of the power of this method remains open. It would be interesting to compare its power to the power of methods based on finer partitions as in \cite{Hellera}.


\begin{thebibliography}{}
\bibitem[Bergsma and Dassios, 2014]{Bergsma}
Bergsma W. and Dassios A. (2014).
\newblock A consistent test of independence based on a sign
covariance related to Kendall’s tau.
\newblock {\em Bernoulli}, 20(2):1006–1028.
\bibitem[Heller et~al., 2013]{Hellerb}
Heller, R., Heller, Y., and Gorfine, M. (2013).
\newblock A consistent multivariate test of association based on ranks of distances
\newblock {\em Biometrika}, 100(2):503-510
\bibitem[Heller et~al., 2016]{Hellera}
Heller, R., Heller, Y., Kaufman, S., Brill, B., and Gorfine M. (2016)
\newblock Consistent distribution-free K-sample and independence
tests for univariate random variables
\newblock{\em Journal of Machine Learning Research} 17(29):1-54
\bibitem[Hoeffding, 1948]{Hoeffding}
Hoeffding, W.
\newblock A non-parametric test of independence
\newblock {\em The Annals of Mathematical Statistics}, 546–557
\bibitem[Weihs et~al., 2015]{Weihs}
Weihs, L., Drton, M., and Leung, D. (2015).
\newblock Efficient Computation of the Bergsma-Dassios Sign Covariance
\newblock{\em Computational Statistics} 31(1):315-328
\end{thebibliography}
 \end{document}